\documentclass[aps,pra,a4paper,superscriptaddress,longbibliography,reprint]{revtex4-1}

\usepackage{amsmath}
\usepackage{amsfonts}
\usepackage{amssymb}
\usepackage{bm}
\usepackage{bbm}

\usepackage{graphicx}

\usepackage{color}

\newcommand{\eq}[1]{Eq.~(\ref{#1})}

\newcommand{\fig}[1]{Fig.~\ref{#1}}

\begin{document}

\title{Optimizing many-body atomic descriptors for enhanced computational performance
of machine learning based interatomic potentials}

\author{Miguel A. Caro}
\email{mcaroba@gmail.com}
\affiliation{Department of Electrical Engineering and Automation,
Aalto University, 02150, Espoo, Finland}
\affiliation{Department of Applied Physics,
Aalto University, 02150, Espoo, Finland}

\date{July 17, 2019}

\begin{abstract}
We explore different ways to simplify the evaluation of the smooth overlap of atomic 
positions (SOAP) many-body atomic descriptor [Bart\'{o}k \textit{et al}., Phys. Rev. B \textbf{87}, 184115 (2013)].
Our aim is to improve the computational efficiency of SOAP-based similarity kernel
construction. While these improved atomic descriptors can be used for general characterization and
interpolation of atomic properties, their main target application is
accelerated evaluation of machine-learning-based interatomic potentials
within the Gaussian approximation potential (GAP) framework [Bart\'{o}k \textit{et al}., Phys. Rev. Lett. \textbf{104}, 136403 (2010)].
We achieve this objective by expressing the atomic densities in an approximate separable form,
which decouples the radial and angular channels. We then express the elements of the
SOAP descriptor (i.e., the expansion coefficients for the atomic densities) in analytical form
given a particular choice of radial basis set. Finally, we derive recursion formulas for the expansion
coefficients. This new SOAP-based descriptor allows for tenfold speedups compared
to previous implementations, while improving the stability of the radial expansion for distant
atomic neighbors, without degradation of the interpolation power of GAP models.
\end{abstract}

\maketitle

\section{Smooth overlap of atomic positions}

Machine learning (ML) applied to materials modeling has rapidly gained widespread attention
within the computational physics, chemistry and materials science communities due to its
ability to speed up the simulation times for accurate prediction of the
properties of materials. In particular, significant speedups are obtained with respect to
simulation times currently required for atomistic simulation within first-principles approaches,
such as
density-functional theory (DFT). These new ML methodologies also grant access to new
time and length scales in the simulation of interatomic interactions, allowing us to solve
outstanding scientific problems whose study has been previously out of reach~\cite{caro_2018}. Several
ML approaches have arisen in recent years for interpolation of interatomic potential
energy surfaces (PES), most notably based on artificial neural networks and
kernel-based regression techniques~\cite{behler_2007,bartok_2010}. All these approaches feed on two
types of data: 1) the observables to be learned and interpolated (e.g., atomic energies and
forces) which are used during the ML training stage and 2) the structural information
that characterizes atomic environments, known in the ML jargon as ``descriptors'', which are to be used both
during the training stage and when interpolating the PES. Traditional
descriptors used for characterization of PESs with ``classical'' (or ``empirical'') force
fields are bond distances, bond angles, improper/dihedral angles, etc., all of which
involve interactions between two, three or, at most, a handful of particles. However,
to make the most out of the newly available ML infrastructure and learn complex PESs
there is a need for accurate, yet computationally inexpensive, many body
descriptors~\cite{behler_2011,huo_2017}.

The smooth overlap of atomic positions (SOAP) is a recently-introduced approach to encode
atomic environments into a rotationally-invariant representation, given by the SOAP vectors~\cite{bartok_2013}.
These many-body atomic descriptors are designed to provide an accurate measure of similarity between
atomic environments, which can then be fed into kernel-based ML algorithms. In
particular, when used in combination with the Gaussian approximation potential (GAP) formalism~\cite{bartok_2010},
SOAP enables accurate and efficient interpolation of potential energy surfaces~\cite{deringer_2017}.
This accuracy has enabled molecular dynamics (MD) simulations of large and complex systems that were previously
out of reach~\cite{caro_2018}. However, compared to
analytical force fields, SOAP-based GAPs are still CPU-expensive, with the evaluation of
SOAP descriptors being the computational bottleneck. Being able to speed up SOAP evaluation would therefore
provide an invaluable tool for making larger system sizes and simulation times accessible to ML-based MD
simulation codes.

The (full) representation of the atomic density underlying the SOAP approach is done
using 3D Gaussians centered at the atomic (nuclear) sites. The atomic density within a cutoff
sphere $S_i(r_\text{cut})$ surrounding and centered on atom $i$ is therefore given by:
\begin{align}
\rho^{(i)} (\textbf{r}) = \sum_{j \in S_i (r_\text{cut})} \rho_j^{(i)} (\textbf{r}),
\end{align}
where the sum extends over all atoms $j$ inside the cutoff sphere, possibly including $i$
itself. A rotationally-invariant comparison of two such densities is achieved by computing their
overlap integral and averaging over all
possible rotations $\hat{R}$ of one of the atomic environments~\cite{bartok_2013}:
\begin{align}
k^\text{SOAP}(i,j) \propto \int \text{d}\hat{R} \left| \int \text{d}\textbf{r} \,
\rho_i^*(\textbf{r}) \rho_j(\hat{R}\textbf{r}) \right|^2,
\label{03}
\end{align}
where the SOAP kernel $k^\text{SOAP}(i,j)$ gives a bounded measure of similarity between the atomic environment
of atom $i$ and the atomic environment of atom $j$. This similarity measure varies between 0 (the
environments are nothing alike) and 1 (the environments are identical). Note that the exponent to which the
overlap integral is raised, 2 in this case, must be greater than one for the SOAP kernel to retain angular
information~\cite{bartok_2013}. Explicitly computing
this integral is impractical from a computational efficiency standpoint, and thus the usefulness of SOAP
is built on the reformulation of this problem. In this context, a discrete representation of the densities is
achieved by expanding them in a basis. Using a combination of radial basis and spherical harmonics allows us to construct a
rotationally-invariant descriptor in vector form, whose components are products of the expansion
coefficients, without the need to explicitly perform the rotation. The expanded density
takes the following form:
\begin{align}
\rho (\textbf{r}) = \sum_{j \in S (r_\text{cut})} \sum_{nlm} c^j_{nlm} \, g_n (r) \, Y_{lm}(\theta, \phi),
\end{align}
where we have omitted the $(i)$ index for compactness. The $\{g_n\}$ is an orthonormal radial
basis and $Y_{lm}(\theta, \phi)$ are the spherical harmonics.

From these expansion coefficients and after some algebraic manipulation, it can be shown that
the SOAP kernel can be expressed as~\cite{bartok_2013}:
\begin{align}
k^\text{SOAP}(i,j) \propto & \sum_{nn'lmm'} c_{nlm}^i (c_{n'lm}^i)^* c_{n'lm'}^j (c_{nlm'}^j)^*
\nonumber \\
= & \sum_{nn'l} p_{nn'l}(i) p_{nn'l}(j), 
\end{align}
where
\begin{align}
p_{nn'l} (i)= \sum_m c^i_{nlm} (c^i_{n'lm})^*
\end{align}
is the \textit{power spectrum} of the atomic density. The vectors given by $\textbf{p} \equiv \{ p_{nn'l} \}$
define, after normalization, the SOAP many-body atomic descriptors:
\begin{align}
\textbf{q}^\text{SOAP} (i) = \frac{\textbf{p}(i)}{\sqrt{\textbf{p}(i) \cdot \textbf{p}(i)}}.
\end{align}
The expression for the SOAP kernel follows in the form of a dot product:
\begin{align}
k^\text{SOAP}(i,j) = \left( \textbf{q}^\text{SOAP} (i) \cdot \textbf{q}^\text{SOAP} (j) \right)^\zeta,
\end{align}
where $\zeta$ is some positive number, usually greater than 1, that controls the ``sharpness'' of the
kernel, that is, the ability of the kernel to emphasize differences between atomic environments (the
larger $\zeta$ the sharper the kernel)~\cite{bartok_2015}.
These SOAP descriptors simplify immensely the task of evaluating \eq{03}. However, there is a number
of further simplifications and modifications of the original SOAP formulation that can dramatically
increase the computational performance of this approach. In the next section
we propose a new form of a SOAP-like many-body descriptor and prove its superior suitability for
computational evaluation.

\section{New SOAP based on pseudogaussian functions}

\begin{figure}[t]
\includegraphics[width=\columnwidth]{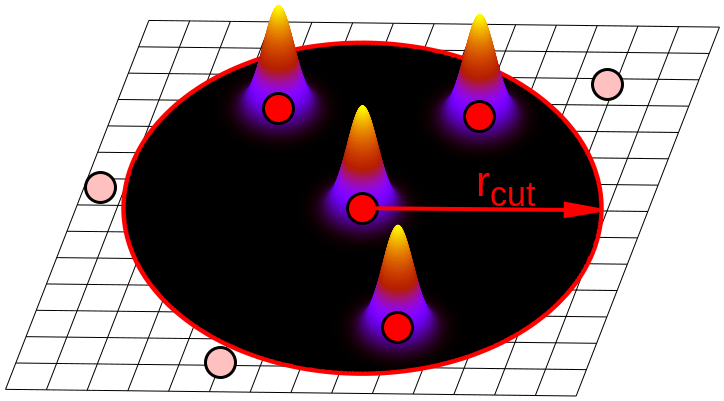}
\caption{Within the SOAP formalism, the atomic neighborhood of an atom is represented,
inside a cutoff sphere, by an atomic density field centered on said atom. The neighbor
atoms beyond the cutoff radius are not taken into account by the SOAP descriptor.}
\label{11}
\end{figure}

In the original SOAP formulation~\cite{bartok_2013}, the atomic densities are represented by atom-centered Gaussian functions (\fig{11}):
\begin{align}
\rho_j (\textbf{r}) = \exp{\left(-\frac{|\textbf{r} - \textbf{r}_j|^2}{2 \sigma^2} \right)},
\end{align}
and the corresponding expansion coefficients $c^j_{nlm}$ have the form:
\begin{align}
c^j_{nlm} = b_{nl}^j \, c^j_{lm}.
\end{align}
The angular dependence of the $b$s arises because, if we were to retain the radial dependence of the
expansion coefficients inside the $c^j_{lm}$, these would take the form~\cite{bartok_2013}:
\begin{align}
c^j_{lm} (r) \equiv 4 \pi \exp{\left[ - \alpha \left( r^2 + r_j^2 \right) \right]}
i_l (2 \alpha r r_j) Y^*_{lm} (\theta_j, \phi_j),
\end{align}
where $\alpha = 1/(2\sigma^2)$ and all the quantities with subindex $j$ refer to the relative position
of atom $j$ with respect to central atom $i$.
The modified spherical Bessel function of the first kind $i_l$, that depends on $r$, introduces the
simultaneous $l$ and $n$ dependence of the coefficients, and at the same time makes their analytical
derivation non-trivial (although still possible, see Refs.~\cite{jager_2018,himanen_2019}).
All the details are given in the original SOAP paper~\cite{bartok_2013}.

To simplify this, we suggest to replace one problem by another. Let us express the atomic density in an approximate \textit{separable} form:
\begin{align}
\rho_j (\textbf{r}) \approx \rho_{r,j}(r) \rho_{\perp,j}(\theta,\phi).
\end{align}
Conveniently, we will continue to use Gaussians:
\begin{align}
\rho_j (\textbf{r}) = \exp{\left[ - \frac{1}{2} \frac{(r - r_j)^2}{\sigma_r^2} \right]}
\exp{\left[ - \frac{1}{2} \frac{{r_{\perp, j}}^2}{\sigma_\perp^2} \right]},
\end{align}
which would be the exact expression of a 3D Gaussian if $r$ was a regular Cartesian dimension
and $r_{\perp, j}$ measured the distances from $\textbf{r}_j$ in the plane perpendicular to $r$ which
contains $\textbf{r}_j$. In our approximation, $r_{\perp, j}$ measures distances from $\textbf{r}_j$
in the spherical cap of radius $r_j$. How well this spherical cap can be approximated by a plane
depends on the ratio $r_{\perp, j} / r_j$, which in practice depends on $\sigma / r_j$, since $\sigma$
controls the decay length of our density as we move away from $\textbf{r}_j$.
The approximation improves as one moves further away from the origin. Therefore, in practice,
we are not modeling our atomic density with Gaussians which are spherically-symmetric about
$\textbf{r}_j$, but about the origin. However, we must stress that this can also be understood as a
choice, rather than an approximation, since in principle we have freedom in how we represent the
atomic density, as long as permutational, translational and rotational invariances are preserved.
An additional advantage of our approach is that we can choose different $\sigma$s for
the radial and angular channels, $\sigma_r$ and $\sigma_\perp$, respectively. This further choice has
the advantage that it reflects on the fact that length-preserving and angle-preserving interatomic
interactions have different characteristic strengths. A final improvement in the choice of $\sigma$ is
to incorporate a radial dependence, as already proposed in Ref.~\cite{willatt_2019} and
directly applicable to the original SOAP descriptor without the modifications incorporated
here.
This radial dependence allows for increasingly ``blurry'' atomic environments
as one moves away from the center of the SOAP sphere. When done together with $r$-dependent downscaling of the
contribution of distant atomic neighbors to the density field, this strategy provides us with a flexible
many-body kernel that is able to precisely encode the structural atomic information required for
accurate interpolation of potential-energy surfaces.

We can approximate $r_{\perp, j}$ as
\begin{align}
r_{\perp, j}^2 \approx & 2 \left( r_j^2 - \textbf{r}_j \cdot r_j \hat{\textbf{r}} \right)
= 2 \left( r_j^2 - r_j^2 (\hat{\textbf{r}}_j \cdot \hat{\textbf{r}}) \right).
\\
= & 2 \left( r_j^2 - r_j^2 \cos{\epsilon_j} \right),
\label{12}
\end{align}
where $\epsilon_j$ is the angle between $\textbf{r}$ and $\textbf{r}_j$. Equation~(\ref{12})
is equivalent to a second-order truncation of the Taylor expansion of the cosine, i.e.,
$\cos{\epsilon_j} \approx 1 - \epsilon_j^2/2$.
With this approximation, the atomic density of atom $j$ is expressed as
\begin{align}
\rho_j (\textbf{r}) =
\underbrace{\exp{\left[ - \frac{(r - r_j)^2}{2 \sigma_{r,j}^2} \right]}}_{(1)}
\underbrace{\exp{\left[ - \frac{r_j^2}{\sigma_{\perp,j}^2} \right]}}_{(2)}
\underbrace{\exp{\left[ \frac{\textbf{r}_j \cdot r_j \hat{\textbf{r}}}{\sigma_{\perp,j}^2} \right]}}_{(3)},
\label{05}
\end{align}
and the total density to be expanded is
\begin{align}
\rho(\textbf{r}) = \sum_j A_j f(r; r_j, r_\text{cut}) \rho_j (\textbf{r} ),
\end{align}
where we have explicitly introduced the amplitude $A_j$ to downscale the contribution
of distant neighbors. In \eq{05} we have notated the $\sigma$s with $j$ subindices
for the same reason. Radial downscaling has been discussed in more detail in Ref.~\cite{willatt_2018}.
Casting the problem in an explicitly separable form means that expansion of the radius-dependent
part of \eq{05} becomes effectively a 1D problem. Therefore, the $r^2$ term that usually accompanies 3D integrals in
spherical coordinates, which originates from the angular line elements, will not appear in our integrals
for radial expansion [cf. \eq{01} and next section].
The function $f(r; r_j, r_\text{cut})$
can be any smoothing function that goes smoothly to zero at the cutoff. The most straightforward approach to downscaling
distant neighbors is to introduce simple radial dependencies for these SOAP hyperparameters:
\begin{align}
\sigma_{r,j} & = \sigma_{0, r} + \alpha_r r_j,  \qquad  \sigma_{\perp,j} = \sigma_{0,\perp} + \alpha_\perp r_j,
\\
A_j &= \frac{1}{\sigma_{r,j} \sigma_{\perp,j}^2} \left( 1 + 2\left(\frac{r_j}{r_\text{cut}}\right)^3
- 3\left(\frac{r_j}{r_\text{cut}}\right)^2 \right)^a,
\label{15}
\end{align}
where $\sigma_{0,r}$ and $\sigma_{0,\perp}$ define the Gaussian representation of the central atom in
the SOAP sphere (although the value of $\sigma_{r,\perp}$ does not really have an effect on the representation
of the central atom, since $r_j = 0$ implies the angular Gaussian equals one always, cf. \eq{12}),
and $\alpha_r, \alpha_\perp, a \ge 0$ (when $\alpha_r = \alpha_\perp = a = 0$ we retrieve the no
downscaling limit). The third-order polynomial introduced in \eq{15} is the simplest function that,
for $a>0$, goes smoothly from 1 at the origin to 0 at the cutoff. For $a \ge 1$, its derivative at
$r_\text{cut}$ is also smooth.

In \eq{05}, (1) is the radial part, (2) is a constant factor and (3) only depends on the angle
between $\textbf{r}$ and $\textbf{r}_j$. Term (3) can be expressed as~\cite{kaufmann_1989}:
\begin{align}
\exp{\left[ \frac{\textbf{r}_j \cdot r_j \hat{\textbf{r}}}{\sigma_{\perp,j}^2} \right]} =
\sum_{l = 0}^{\infty} (2l+1) \, i_l \left( \frac{r_j^2}{\sigma_{\perp,j}^2} \right) P_l (\cos{\epsilon_j}),
\end{align}
where $P_l$ is the Legendre polynomial. The addition theorem allows us to express this as~\cite{kaufmann_1989}:
\begin{align}
4 \pi \sum_{l=0}^\infty i_l \left( \frac{r_j^2}{\sigma_{\perp,j}^2} \right) \sum_{m=-l}^l Y_{lm}^*(\theta_j, \phi_j) Y_{lm}(\theta, \phi).
\end{align}

We can obtain our radial expansion coefficients as
\begin{align}
b_n^j = \int_0^{r_\text{cut}} \text{d}r \, g_n(r) A_j f(r; r_j, r_\text{cut}) \exp{\left[ - \frac{1}{2} \frac{(r - r_j)^2}{\sigma_{r,j}^2} \right]},
\label{01}
\end{align}
where $r_\text{cut}$ is the SOAP sphere cutoff radius.
With a polynomial basis (or a number of bases, for that matter) the $b$s have analytical form, as shown in the
next section.
Our final coefficients are:
\begin{align}
c_{nlm}^j = 4 \pi \, b_n^j \, \exp{\left( - \frac{r_j^2}{\sigma_{\perp,j}^2}
\right)} (2l+1) \, i_l \left( \frac{r_j^2}{\sigma_{\perp,j}^2} \right) Y_{lm}^*(\theta_j, \phi_j).
\label{08}
\end{align}
Given that the $b$s have an analytical form, all of these coefficients can be obtained analytically.
Furthermore, even though the product $i_l(x^2) \exp{\left(-x^2\right)}$ can be numerically unstable for large $x$
when the two factors are computed independently and then multiplied, when we compute the combined function
the product is quite stable. For small $x$ the function is divergent but it can be computed as a limit
using the Taylor expansion of $\exp{\left(-x^2\right)}$.

The total coefficients are obtained by summing over atomic contributions:
\begin{align}
c_{nlm} = \sum_j c_{nlm}^j,
\end{align}
and, from them, the power spectrum is given by
\begin{align}
p_{n n' l} = \sum_{m = -l}^l c_{nlm} c_{n' l m}^*.
\end{align}
The $c_{nlm}$ are symmetric with respect to $m$, because of the properties of the
spherical harmonics: $c_{nl-m} = (-1)^m c_{nlm}^*$. We can thus simplify the evaluation of $p_{nn'l}$ as follows:
\begin{align}
\sum_{m = -l}^l c_{nlm} c_{n' l m}^* = & c_{nl0} c_{n' l 0}^* + \sum_{m=1}^{l} \left( c_{nlm} c_{n' l m}^* + c_{n'lm} c_{n l m}^* \right)
\nonumber \\
& c_{nl0} c_{n' l 0}^* + 2 \sum_{m=1}^{l} \text{Re}\left\{ c_{nlm} c_{n' l m}^* \right\},
\label{09}
\end{align}
which reduces the number of terms in the sum considerably, and also reduces the number of coefficients which need
to be computed, since for $m<0$ the corresponding $c_{nlm}$ does not need to be computed (because
it is not used). It also follows from \eq{09} that the $p_{n n' l}$ are symmetric upon exchange of $n$ and $n'$,
which reduces the number of evaluations even further.

\vfil

\section{Recursion formulas for the expansion coefficients}

\subsection{Radial expansion coefficients}

The $b$s can be computed analytically for certain bases. In the original SOAP paper~\cite{bartok_2013},
3rd- and higher-order polynomials were proposed, even though they were not implemented in practice. Here
we will use these polynomials since they allow us to infer recursion formulas for the radial expansion
coefficients, as will be shown later.
If our polynomial basis is $\{ \phi_\alpha (r) \}$, the orthonormal basis $\{ g_n (r) \}$ is constructed as follows:
\begin{align}
g_n (r) = \sum_{\alpha = 1}^{n_\text{max}} W_{n\alpha} \phi_\alpha(r),
\end{align}
with
\begin{align}
\phi_\alpha (r) = \left( 1 - \frac{r}{r_\text{cut}} \right)^{\alpha+2} / N_\alpha,
\end{align}
where $N_\alpha = \sqrt{r_\text{cut} / (2\alpha + 5)}$ is a normalization factor.
These polynomials, and their first and second derivatives,
conveniently go to zero at the cutoff. The $W_{n\alpha}$ are obtained from the overlap
matrix $S$ and need to be obtained only once for a given $n_\text{max}$ (they could even be tabulated):
\begin{align}
W = S^{-1/2},
\\
S_{\alpha \beta} = \int_0^{r_\text{cut}} \text{d}r \phi_\alpha(r) \phi_\beta(r).
\end{align}
In practice, we use these polynomials for $\alpha = 1, \dots, n_\text{max}-1$ and augment
our basis set with a Gaussian function centered at the origin,
which allows us to resolve the central atom in the atomic environment exactly:
\begin{align}
\phi_{n_\text{max}} (r) = \frac{\sqrt{2}}{\sqrt{ \sigma_{0,r}}\pi^{1/4}} \exp{\left[ - \frac{r^2}{2 \sigma_{0,r}^2} \right]}.
\label{06}
\end{align}
The normalization factor for this auxiliary Gaussian basis function assumes that $r_\text{cut} \gg \sigma_{0,r}$.

When computing the radial expansion coefficients, we do not evaluate \eq{01}, i.e., the $b_n^j$. Instead, we compute
the overlap integrals for the $b_\alpha^j$:
\begin{align}
b_\alpha^j = \int_0^{r_\text{cut}} \text{d}r \, \phi_\alpha(r) \exp{\left[ - \frac{1}{2} \frac{(r - r_j)^2}{\sigma_{r,j}^2} \right]}.
\label{19}
\end{align}
We have left the $A_j f(r; r_j, r_\text{cut})$ term [cf. \eq{01}]
out of \eq{19} for simplicity, but without loss of generality since, as we will show later on, for our particular 
choice of smoothing function the functional form of the atomic density remains unchanged.
From these $b_\alpha^j$, the transformation is straightforward:
\begin{align}
b_n^j = \sum_{\alpha = 1}^{n_\text{max}} W_{n\alpha} b_\alpha^j.
\end{align}
The reason for working with the $b_\alpha^j$ is that the polynomial form of the $\{ \phi_\alpha \}_1^{n_\text{max}-1}$ can be exploited to
derive recursive relations. Consider the following integration by parts (where for clarity we have omitted the
integration limits):
\begin{widetext}
\begin{align}
N_\alpha b_\alpha^j = &
\int \text{d}r \left(1-\frac{r}{r_\text{cut}}\right)^{\alpha+2} \exp{\left[ - \frac{\left(r-r_j\right)^2}{2 \sigma_{r,j}^2}\right]}
\nonumber \\
= & - \left( 1-\frac{r}{r_\text{cut}} \right)^{\alpha+3} \frac{ r_\text{cut}}{\alpha+3} \exp{\left[ - \frac{\left(r-r_j\right)^2}{2 \sigma_{r,j}^2}\right]}
+ \int \text{d}r \left( 1-\frac{r}{r_\text{cut}} \right)^{\alpha+3} \frac{ r_\text{cut}}{\alpha+3} \frac{r_j - r}{\sigma_{r,j}^2} \exp{\left[ - \frac{\left(r-r_j\right)^2}{2 \sigma_{r,j}^2}\right]}.
\label{02}
\end{align}
\end{widetext}
We can manipulate the $r_j - r$ term in \eq{02} as follows:
\begin{align}
r_j - r = r_j - r_\text{cut} + r_\text{cut} \left( 1 - \frac{r}{r_\text{cut}} \right),
\end{align}
so that, after collecting the terms, \eq{02} reads as
\begin{align}
N_\alpha b_\alpha^j = & - \left( 1-\frac{r}{r_\text{cut}} \right)^{\alpha+3} \frac{ r_\text{cut}}{\alpha+3} \exp{\left[ - \frac{\left(r-r_j\right)^2}{2 \sigma_{r,j}^2}\right]}
\nonumber \\
& + \frac{r_\text{cut} \left( r_j - r_\text{cut} \right)}{\left( \alpha + 3 \right) \sigma_{r,j}^2} N_{\alpha+1} b_{\alpha+1}^j
\nonumber \\
& + \frac{r_\text{cut}^2}{\left( \alpha + 3 \right) \sigma_{r,j}^2} N_{\alpha+2} b_{\alpha+2}^j.
\end{align}
This recursion formula can be rewritten as
\begin{align}
b_{\alpha}^j = &\left( 1 - \frac{r}{r_\text{cut}} \right)^{\alpha+1} \frac{\sigma_{r,j}^2}{r_\text{cut} N_\alpha} \exp{\left[ - \frac{\left(r-r_j\right)^2}{2 \sigma_{r,j}^2}\right]}
\nonumber \\
& + \frac{N_{\alpha-1}}{N_{\alpha}} \left(1 - \frac{r_j}{r_\text{cut}} \right) b_{\alpha-1}^j 
+ \frac{N_{\alpha-2}}{N_{\alpha}} \frac{\left( \alpha + 1 \right) \sigma_{r,j}^2}{r_\text{cut}^2} b_{\alpha-2}^j.
\label{04}
\end{align}
While it would appear that we need $b_{-1}^j$ and $b_{0}^j$ to obtain the first coefficient, it can be shown that we
can start the sequence by obtaining $b_{-1}^j$ from $b_{-2}^j$ assuming $b_{-3}^j = 0$. Therefore,
we only need
\begin{align}
b_{-2}^j = \frac{\sigma_{r,j}}{N_{-2}} \sqrt{\frac{\pi}{2}} \left. \text{erf}{\left[ \frac{r-r_j}{\sqrt{2}\sigma_{r,j}} \right]} \right|_0^{r_\text{cut}},
\end{align}
and the rest of the radial expansion coefficients are obtained from the recursion relation, \eq{04}.

The recursion formulas above are valid for $\alpha < n_\text{max}$. The overlap integral between the atom's
Gaussian and the auxiliary Gaussian function $\phi_{n_\text{max}}(r)$ given by \eq{06} is
straightforward to derive:
\begin{align}
&\int\limits_0^{\infty} \text{d}r \, \phi_{n_\text{max}}(r) \exp{\left[ - \frac{\left(r-r_j\right)^2}{2 \sigma_{r,j}^2}\right]} =
\nonumber \\
&\frac{\pi^{1/4}}{\sqrt{\sigma_{0,r}}} \exp{\left[ - \frac{r_j^2}{2 \sigma_*^2} \right]}
\frac{\sigma_{0,r} \sigma_{r,j}}{\sigma_*} \left( 1 + \text{erf}{\left[\frac{\sigma_{0,r} r_j}{\sqrt{2}\sigma_{r,j} \sigma_*} \right]} \right),
\end{align}
where $\sigma_* = \sqrt{\sigma_{0,r}^2 + \sigma_{r,j}^2}$. We have assumed above that the integrand, that is,
the overlap between $\phi_{n_\text{max}}(r)$ and $\rho_j(r)$, has effectively decayed to zero at the cutoff.

Finally, we must remark that with our choice of radial basis, linear dependencies in the overlap matrix $S$ develop
for $n_\text{max} > 12$, i.e., a singular-value decomposition of $S$ yields some very small eigenvalues. Therefore,
our implementation becomes unstable for $n_\text{max} > 12$.

\subsection{Density smoothing at the cutoff}

\begin{figure}[t]
\includegraphics[]{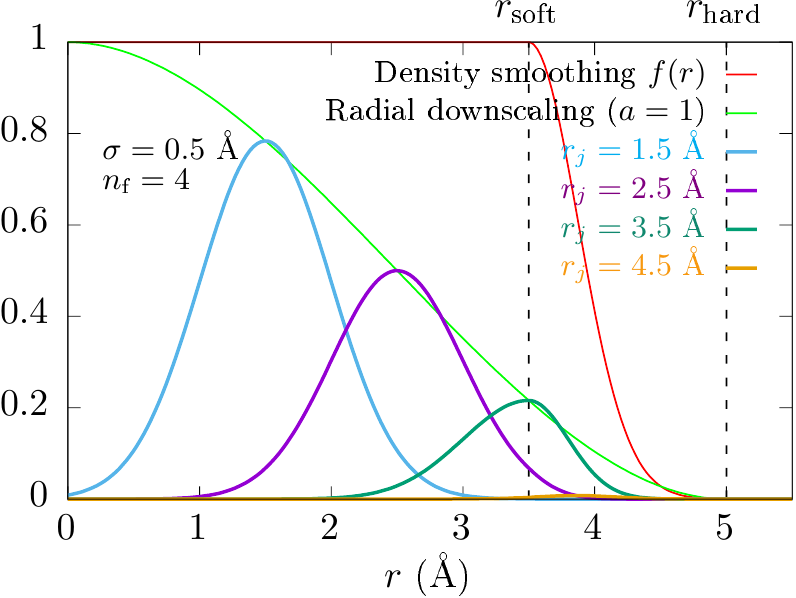}
\caption{Example of smoothed radial atomic densities for different values of $r_j$.}
\label{07}
\end{figure}

To ensure smoothness of interpolated potential energy surfaces and any general ML model based on SOAP,
it is vital to remove sharp discontinuities of the kernel functions when atoms move in and out of the
cutoff sphere~\cite{bartok_2013}. In our present implementation, we rely on a soft cutoff, a hard cutoff and
a ``buffer zone'' in between. The hard cutoff delimits the sphere within which the SOAP descriptor
``sees'' neighboring atoms: any atom outside of the hard cutoff will be completely neglected. The soft
cutoff delimits the sphere within which the atomic densities are represented fully, as per the
expressions given above. The buffer zone is the region between soft and hard cutoff within which the
atomic densities are smoothed out to zero. Therefore, a suitable smoothing function is defined as:
\begin{align}
f(r; r_\text{soft}, r_\text{hard}) =
\begin{cases}
1 & \text{ if } r<r_\text{soft},
\\
0 & \text{ if } r>r_\text{hard},
\end{cases}
\end{align}
and a smooth transition from 1 to 0 in all other cases. Note that this smoothing affects the \textit{density}
field; we have already introduced another smoothing function in \eq{15} to downscale the heights of the
Gaussians. We choose the following convenient definition for our density smoothing function:
\begin{align}
f(r) =
\begin{cases}
1 & \text{ if } r < r_\text{soft},
\\
\exp{\left[ -\frac{n_\text{f}^2}{2} \frac{\left(r - r_\text{soft}\right)^2}{\left( r_\text{hard} - r_\text{soft}
\right)^2}\right]} & \text{ if } r_\text{soft} \le r \le r_\text{hard},
\\
0 & \text{ if } r > r_\text{hard}.
\end{cases}
\end{align}
The characteristic decay length is selected by choosing a suitable filtering parameter $n_\text{f}$, such
that, numerically, the exponential is approximately zero at $r_\text{hard}$. For instance,
$n_\text{f} = 4$ already brings the smoothing function down to $\sim 3.35 \times 10^{-4}$ at the hard cutoff,
regardless of the actual choice of cutoffs. Filtering parameters equal and larger than 4 are suitable choices,
noting that choosing a very large number actually defeats the purpose of using a buffer zone.
Therefore our implementation uses $n_\text{f} = 4$ as default. The motivation for using a Gaussian
as smoothing function is simple: since the product of two Gaussians (the atomic density and the smoothing
function) is also a Gaussian, we can use the same recursion relations derived in the previous section,
choosing the integration limits appropriately. That is, the expansion is divided into the $[0, r_\text{soft}]$
and $[r_\text{soft}, r_\text{hard}]$ domains. Within the first domain, we expand $\rho_j (r)$, whereas
within the second domain we expand $f(r) \rho_j (r)$, where $f(r) \rho_j (r)$ is also a Gaussian. In both
cases, the overlap integrals are scaled by the downscaling factor introduced in \eq{15}. Figure~\ref{07}
shows an example of how the smoothing procedure outlined above works in practice.

\subsection{Angular expansion coefficients}

Compared to the radial expansion coefficients, obtaining the angular part of \eq{08}, that is everything that depends
on $l$ and/or $m$, may seem trivial. However,
there is still a number of simplifications that can be implemented in order to optimize computational performance.
We start out with the product of the exponential and Bessel functions in that equation, for which we define
the ilexp function:
\begin{align}
\text{ilexp} (x;l) \equiv e^{- x^2}  i_l \left( x^2 \right).
\end{align}
Based on the recursion relation for the modified spherical Bessel function of the first kind $i_l$~\cite{arfken_1999},
we can derive the following recursion relation for $\text{ilexp}(x;l)$:
\begin{align}
\text{ilexp}(x;l) = \text{ilexp}(x;l-2) - \frac{(2l-1)}{x^2} \text{ilexp}(x;l-1),
\end{align}
for which we need the first two functions to start the sequence:
\begin{align}
&\text{ilexp}(x;0) = \frac{1 - e^{-2 x^2}}{2 x^2},
\nonumber \\
&\text{ilexp}(x;1) = \frac{x^2 - 1 + e^{-2 x^2} \left( x^2 + 1\right)}{2 x^4}.
\end{align}
For $x$ close to zero, we use the Taylor expansion of the exponential part to avoid the singularity:
\begin{align}
&\text{ilexp}(x; 0) = 1 - x^2
\nonumber \\
&\text{ilexp}(x; l) = \frac{x^{2l}}{(2l+1)!!}.
\end{align}
These expressions and recursion relations allow us to, computationally, obtain all the $\text{ilexp}(x;l)$
functions, from $l = 0$ to $l = l_\text{max}$, for the same cost of obtaining $\text{ilexp}(x;l_\text{max})$.

The next step for the angular expansion is to optimize the evaluation of the (complex conjugate) of the
spherical harmonics $Y_{lm}^*(\theta_j, \phi_j)$:
\begin{align}
Y_{lm}^*(\theta_j, \phi_j) = \sqrt{\frac{\left( 2l + 1\right ) \left( l -m \right)!}{4 \pi \left( l +m \right)!}}
e^{-\text{i}m\phi_j} P_{lm} \left( \cos{\theta_j} \right).
\end{align}
Computationally, evaluating this equation can be divided into three tasks: i) computing the prefactors, ii)
computing the complex exponentials and iii) computing the Legendre polynomials. The first simplification is to
obtain only the terms for which $m \ge 0$, cf. \eq{09}. After that, a second simplification is that all of these
three tasks can be expressed as a recursion series. The calculation of the factorial terms is trivially recursive,
by varying $m$ for fixed $l$. The calculation of the complex exponential is, perhaps surprisingly, rather
expensive computationally if implemented naively:
\begin{align}
e^{-\text{i}m\phi} = \cos \left( m \phi \right) - \text{i} \sin \left( m \phi \right),
\end{align}
where we have used Euler's formula. Modern compilers will take a significant amount of time evaluating
these trigonometric functions. Instead, we can use Chebyshev's recursion formula to considerably speed up this evaluation:
\begin{align}
&\cos \left( m \phi \right) = 2 \cos \phi \, \cos \left[ \left(m-1\right) \phi \right] - \cos \left[ \left(m-2\right) \phi \right],
\nonumber \\
&\sin \left( m \phi \right) = 2 \cos \phi \, \sin \left[ \left(m-1\right) \phi \right] - \sin \left[ \left(m-2\right) \phi \right],
\end{align}
where we only need to call the compiler's implementation of the intrinsic functions $\cos$ and $\sin$ twice: once
for $\cos (-\phi) = \cos \phi$ and once for $\sin (-\phi) = - \sin\phi$. All the other function
calls, up to $m = l_\text{max}$, are to sums and multiplications, which are significantly faster.

Finally, the calculation of associated Legendre polynomials can also be cast as a (rather more complicated)
recursion. We need the following six polynomials to initialize the recursion series:
\begin{align}
\begin{array}{ll}
P_{00}(x) = 1,
&
P_{10}(x) = x,
\\
P_{11}(x) = - \sqrt{1 - x^2},
&
P_{20}(x) = \frac{3}{2} x^2 - \frac{1}{2},
\\
P_{21}(x) = -3 x \sqrt{1 - x^2},
&
P_{22}(x) = 3 - 3 x^2.
\end{array}
\end{align}
With these, we first need to obtain $P_{l0}$ and $P_{l1}$ with a recursion formula on $l$:
\begin{align}
P_{lm}(x) = \frac{(2l-1) x P_{l-1 \, m}(x) - (l-1+m) P_{l-2 \, m}(x)}{l-m}.
\end{align}
From those, we now get all the $P_{lm}$ for greater $m$ using a recursion formula on $m$:
\begin{align}
P_{lm}(x) = & - \frac{2(m-1)x}{\sqrt{1-x^2}} P_{l \, m-1}(x) 
\nonumber \\
& - (l+m-1)(l-m+2)P_{l \, m-2}(x).
\end{align}
For values of $x$ very close to $\pm 1$ these recursion formulas diverge, even though the actual $P_{lm}$ are finite.
In that case we simply set all the $P_{lm}$, for $l \ge 3$ and $m \ge 2$, to zero. In our current implementation
we establish the condition $||x|-1| < 10^{-5}$ to consider $x$ to be ``very close'' to $\pm 1$.

\subsection{Speed}

We tested our new implementation, written in Fortran, for speedup with respect to the implementation available from the QUIP code
(also written in Fortran)~\cite{ref_quip}. QUIP is linked through the interface available from Quippy. Since not only the algorithms to
compute the SOAP descriptors but also the software implementations differ (even though they both use the same language),
we attempt as fair a comparison as possible by only timing the time it takes both codes to carry out the atomic density
expansion and construction of the SOAP vectors, that is, excluding extraneous operations like nearest-neighbor list builds
and such. Our test system is an atomic structure made out of 10000 atoms randomly placed within a cubic box of side length
$L = 46.371$~{\AA} and constrained to be not closer than 0.7~{\AA} from one another.

\begin{figure}[t]
\includegraphics[]{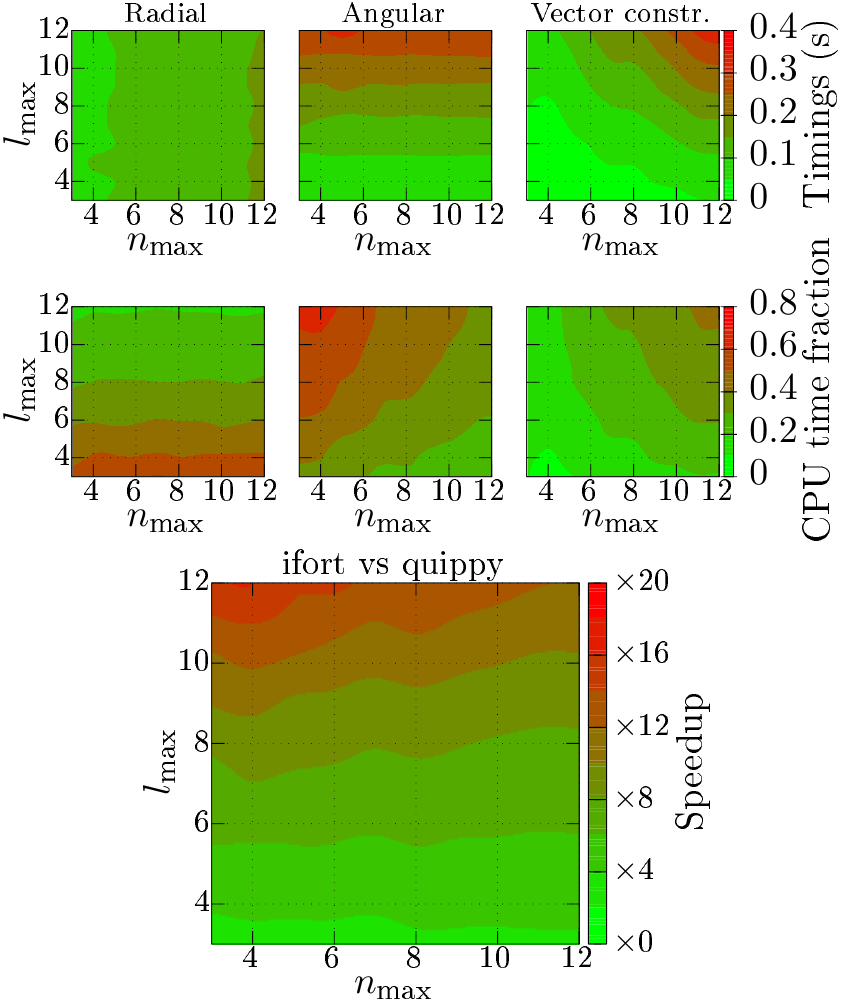}
\caption{Timings (on a single core) for the different steps, i.e., radial expansion, angular expansion, and SOAP vector construction (from
the expansion coefficients) of our current implementation.
``CPU time fraction'' gives the portion of the total execution time that each task
takes to run (at any given combination of $n_\text{max}$ and $l_\text{max}$ the sum for
the three tasks equals one).
Speedup refers to the inverse ratio of the timing of
our new SOAP computation compiled with ifort compared to the Quippy implementation ($t_\text{quippy} / t_\text{ifort}$).}
\label{10}
\end{figure}

For Quippy, we can only get the overall execution time for density expansion plus SOAP vector construction. For our
implementation, we can get the timing for each of the three steps individually: radial expansion, angular expansion and
vector construction. We tried our implementation built with two different compilers: the proprietary Intel compiler
(``ifort'') and the free ``gfortran'' compiler. Although ifort achieved significantly better performance than gfortran
with ``aggressive'' optimization flags (circa 40\% better) this led to numerical instabilities. With numerically-safe
optimization the ifort binary improves gfortran's binary by approximately 20\%. We only report timings for the ifort-built
binary here. It is possible that by adapting the current code to prevent numerical instabilities better performance
can be achieved with ifort or other proprietary compilers. However, we have not exhaustively attempted this for the present work.
The results of our tests, for all the basis sets that can be constructed by combinations of $3 \le n_\text{max} \le 12$
and $3 \le l_\text{max} \le 12$, are given in \fig{10}.

We start our discussion of performance by stating that basis set sizes typically used to
train and evaluate accurate
GAPs range in the $8 \le n_\text{max} \le 12$ and $8 \le l_\text{max} \le 
12$ intervals~\cite{deringer_2017,deringer_2018b}. We should therefore keep
these ranges in mind when establishing the speedup factors that will be achieved with the new
implementation in practical applications. First, we look at timings for the individual steps, i.e.,
the top row in \fig{10}. Remarkably, we can compute all the SOAP descriptors for our 10000 atoms,
with accurate settings on a single-core CPU, in under one second. Even though computing the radial
expansion coefficients in the small basis set region is the computational bottleneck, the
increasing cost of adding more radial basis functions is quite modest thanks to the recursion
relations, and the angular expansion (whose number of coefficients grows quadratically with
$l_\text{max}$) becomes the bottleneck as the size of the basis grows further. Interestingly,
the cost of SOAP vector construction, that is the multiplication and summation operations on
the individual expansion coefficients that lead to the final SOAP vectors, takes up a significant
fraction of the CPU time in the region of highly accurate representation (middle row in \fig{10}).
In the most typical region of interest, $n_\text{max} = 10$ and $l_\text{max} = 10$, each task
takes about one third of the execution time. Therefore, there is no obvious computational
bottleneck in our current implementation.

The most important panel in \fig{10} is the lower one, where a comparison with
the existing QUIP-SOAP implementation is presented. In our regions of interest, the speedup that we
can achieve is approximately tenfold. As a matter of fact, the practical speedup is even higher
because, as we will show in the next section, the new descriptor is also more accurate and better
able to capture the chemistry of atomic environments. This means that the size of the basis
required to achieve the same model performance with the new SOAP descriptor is actually smaller
than with the old SOAP descriptor.

\section{GAP model performance}

Although the main objective of this paper is to improve the computational efficiency of SOAP
calculations, we need to ensure that the introduced modifications to both atomic density
representation and its basis expansion do not degrade the performance of ML models based on SOAP
kernels. In other words, we need to ensure that the accuracy that can be obtained by a ML model
of interatomic interactions, in terms of average error per atom, that employs the new SOAP is
at least as low as what can be achieved with the original descriptor. We start by training an
interatomic potential for amorphous carbon, followed by an adsorption energy model for the same
material. In both cases we generate the model within the GAP framework, which is the typical
ML-based interatomic potential framework that we expect will make use of the new SOAP.
We also discuss in some detail what is the role of hyperparameters on model
accuracy and how to optimize their values.

\subsection{Cohesive energy model}\label{18}

We trained a ``cohesive energy'' GAP model (i.e., a regular interatomic potential) for amorphous
carbon (a-C) using the database from Deringer and Cs\'anyi~\cite{deringer_2017}. For the purposes of the
current benchmark, our GAP model incorporates only SOAP descriptors, unlike the original a-C GAP that
incorporates also two- and three-body descriptors. Briefly, within the GAP formalism, an interpolated
local atomic energy for environment $i$, $\bar{\epsilon}_i$, is given by a linear combination of kernel
functions:
\begin{align}
\bar{\epsilon}_i = \sum_{s \in S_\text{sparse}} \alpha_s k(i,s),
\end{align}
where the sum runs over environments in the ``sparse'' set (a subset of representative
atomic environments in the training set). The fitting coefficients, $\alpha_s$, are precomputed
during the training stage. The number of coefficients, i.e., the ``size'' of the ML model,
depends on the number of sparse configurations $N_s$, and the cost of evaluating the model grows
linearly with this number. The total number of configurations in the training set $N_t$ is much
larger than $N_s$: while all these $N_t$ configurations are used in deriving the $\alpha_s$
during the training stage, only $N_s$ configurations are used during production calculations. 
A practical guide to train GAPs, including notes on sparsification
of the training set, is given in Ref.~\cite{bartok_2015}. Here, we focus only on the effect of
sparse set size on the performance of GAP models trained using the new SOAP versus the old SOAP.
The a-C database used contains approximately 4k supercells of different sizes with a total of
170k unique local atomic environments, and three times as many forces. Our tests consist of models
trained with $N_s$ values between 100 and 1000, and we trained 10 different models for each value
of $N_s$, where the local environments in the sparse set were chosen randomly. For training, we
can use all $N_t$ local environments and add the corresponding $3N_t$ forces. Currently, our
new SOAP implementation is lacking kernel derivatives and, because of this, we have only trained
old SOAP-based models including forces (this capability is available through QUIP).
We have tested the model performance, computed as the
root-mean square error (RMSE) per atom, on a set of 50 different 64-atom a-C structures that were not
included in the training set.

\begin{figure}[t]
\includegraphics[]{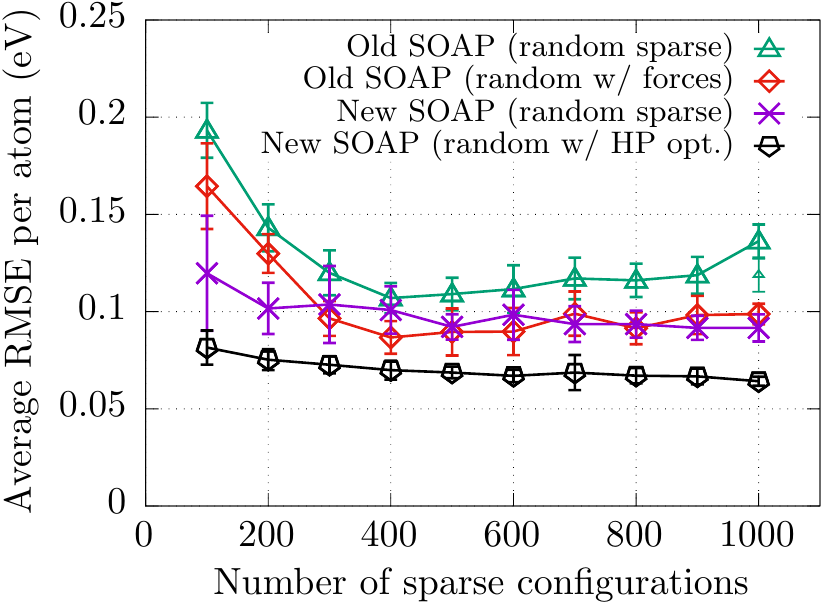}
\caption{Root-mean square error that can be obtained with the old and new SOAP
descriptors/implementations as a function of sparse set size for cohesive energy GAP models.
These GAP models were
trained and tested on a-C data~\cite{deringer_2017}. Optimization of hyperparameters,
as available for the new SOAP, can dramatically improve model performance.
The small symbol for the old SOAP (w/o forces) at $n_\text{sparse} = 1000$ shows how the
fit can be improved at high sampling by increasing the regularization parameter (see text).}
\label{13}
\end{figure}

The results of our test are given in \fig{13}. For the first three models (``old SOAP'' with and
without forces and ``new SOAP'') we choose $n_\text{max} = l_\text{max} = 8$,
$\sigma_{0,r} = \sigma_{0,\perp} = 0.5$~{\AA}, $\zeta=4$ and $r_\text{cut} = 4.5$~{\AA},
very similar to the
parameters used in Ref.~\cite{deringer_2017} to fit the original a-C GAP, and switch off the
extra hyperparameters in the new SOAP that are not available from the old one, to ensure a fair
comparison. For the last model, we
train a ``new SOAP'' with optimized hyperparameters, as discussed in more detail in the next section.
Namely, we choose $n_\text{max} = l_\text{max} = 8$, $\sigma_{0,r} = 0.2$~{\AA},
$\sigma_{0,\perp} = 0.4$~{\AA}, $\alpha_r = 0.08$, $\alpha_\perp = 0.08$,
$a = 1$, $\zeta = 3$ and $r_\text{cut} = 4.5$~{\AA}.

We observe that, with similar sets of hyperparameters, the new SOAP allows us to train an a-C GAP
that is between 10\% and 30\% more accurate than with the old SOAP, depending on the number of
sparse configurations used. Even when forces are added to the fit, the new SOAP (without forces
in the training set) is as accurate as the old SOAP, and more accurate for small sparse sets.
We also note a worsening of the fit for the old SOAP (without forces) as the
number of sparse samples increases. It is possible that, as the number of configurations
in the sparse set is increased beyond the optimum, the extra configurations result in added data noise,
which worsens the fit. This noise can be party removed by adding local information (forces) to the
fit and/or using a less noisy structural kernel (new SOAP). Another strategy to mitigate this problem
is to tune the regularization parameter, by increasing it as the number of sparse configurations go up.
The small symbol at $n_\text{sparse} = 1000$ shows that a regularization parameter twice as large as
the regular one seems to improve the fit in this case.

We conclude that, when used in combination with optimized hyperparameters, using the new SOAP allows us to reduce
the error in the fit by nearly half. We attribute most of this improvement to two factors, one numerical
and one physical. From the numerical perspective, the new SOAP implementation uses improved radial
basis functions which are better at resolving narrower atom Gaussians and work better for longer
cutoff radii. The improvement of the underlying physical model stems from the freedom to choose
radial sigmas and angular sigmas independently. This reflects on the fact that interatomic
interactions (``force constants'' in the context of empirical force fields) have
different characteristic strengths in the angular and radial directions. The separable form of
the new SOAP descriptor allows us to incorporate this physically-motivated effect into the
mathematical representation of the atomic environments.

\subsection{Adsorption energy model and role of hyperparameter on model performance}\label{16}

\begin{figure}[t]
\includegraphics[width=\columnwidth]{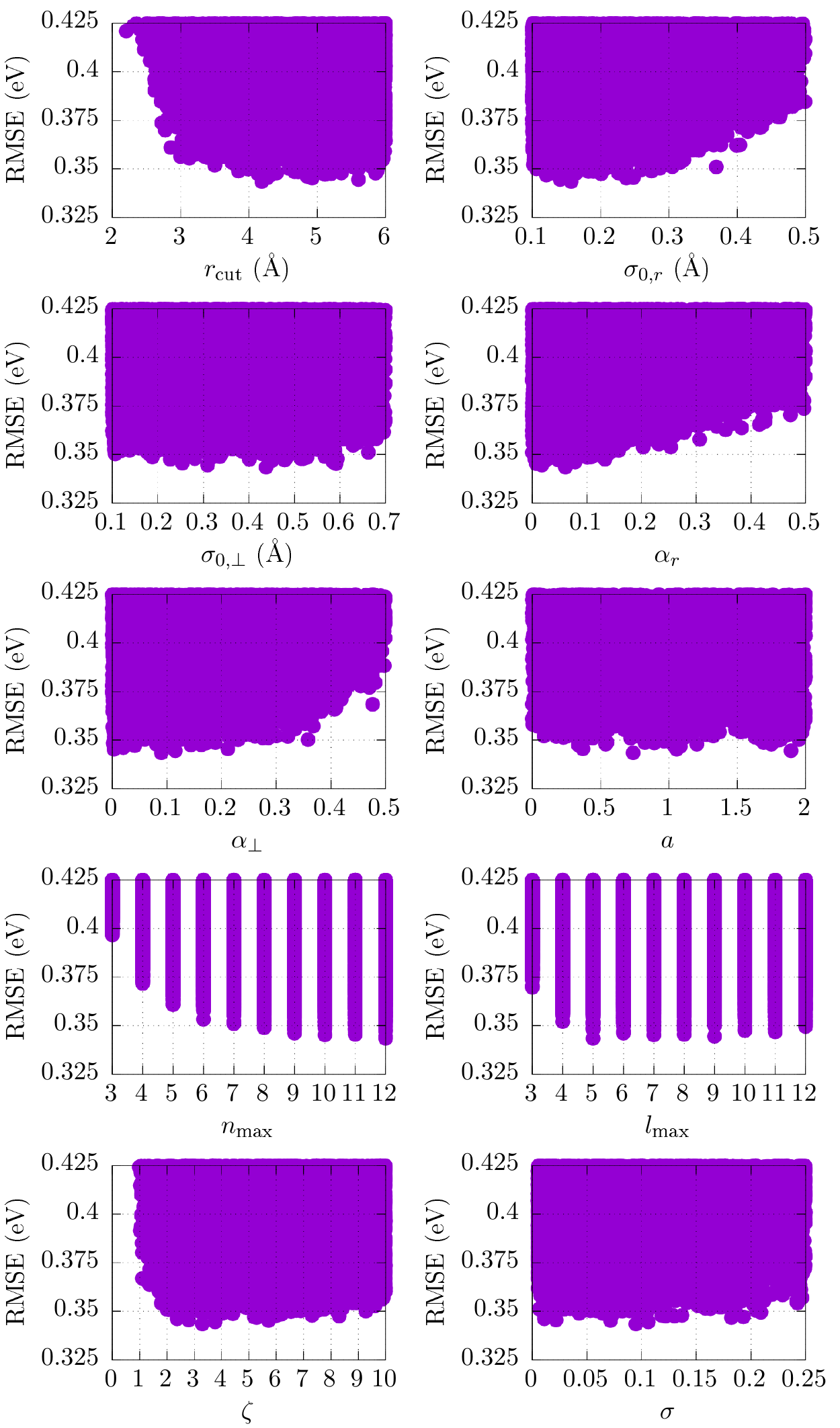}
\caption{Convex hulls for adsorption energy GAP models, trained and tested on data for hydrogen
adsorbed on a-C~\cite{caro_2018c} with the new SOAP descriptor.}
\label{14}
\end{figure}

Training cohesive energy GAP models is computationally expensive, because of the amount of data
involved in the training necessary to obtain a reasonable fit. Therefore, a systematic assessment
of model performance versus choice of hyperparameters (HPs) is impractical over wide regions of
HP space. By contrast, an adsorption energy GAP model is cheap to train. We introduced
such a model for hydrogen adsorption on a-C in our previous work~\cite{caro_2018c} and explored
the idea of HP optimization via Monte Carlo sampling of HP space. While
this optimization method is quite expensive compared to, e.g., Bayesian optimization, it allows us
to ``explore'' wide regions of HP space and get a glimpse of how different parameters affect model
performance. We retrained a large number of adsorption energy models (hundreds of thousands) using
the data from Ref.~\cite{caro_2018c} with our new SOAP descriptor, which allows us to reconstruct
the convex hull for model performance versus HP choice, as shown in \fig{14}. In the figure we can
observe how some parameters have a modest impact on the model performance, such as the regularization
term $\sigma$, while others have a very pronounced effect, e.g., the cutoff radius $r_\text{cut}$.
Compared to the same analysis that we carried out for the same data using the old SOAP
descriptor~\cite{caro_2018c}, we observe that the new SOAP allows us to use the
information of more distant neighbors to improve the accuracy of the model. That is, the performance
of the new SOAP does not degrade significantly as $r_\text{cut}$ is increased beyond the optimum.
At the same time, the data on \fig{14}
clearly show that the optimal values for $\sigma_r$ and $\sigma_\perp$ are different. The choice of
HPs that allowed to obtain a dramatic improvement in cohesive energy GAP accuracy, shown in the
previous section in \fig{13}, was informed from the results obtained for an adsorption energy GAP shown
in \fig{14}. This clearly hints towards the idea of HP transferability across different ML models
that feed on the same kind of atomic information.

\subsection{Hyperparameter optimization for interatomic potentials}

As we have just shown, the particular choice of HPs can dramatically
affect the performance of a ML model for interatomic interactions. While the
adsorption energy model presented in Sec.~\ref{16}
is computationally cheap to train, training \textit{a single} GAP interatomic potential
with typical database sizes, usual sparsification and including forces in the fit can take up
to a few hundred CPU hours. Stochastic evaluation of an error-based objective function, such as
the RMSE, for different combinations of HPs, can thus become a huge computational task.

Comparing to the original SOAP formulation, the SOAP-like descriptor introduced in
this paper incorporates new HPs that make optimization, usually reliant on heuristics, even
more difficult. Therefore, finding efficient ways to obtain combinations of HPs optimally suited to the
problem at hand becomes necessary. A promising route towards improving GAP accuracy via tuning of HPs is
Bayesian optimization~\cite{snoek_2012}. Bayesian optimization models the objective function as a
sample from a Gaussian process compatible with the current set of observations. In our case, an observation
is an evaluation of the RMSE of a GAP for a particular combination of HPs. Employing Bayesian optimization
we can then predict: 1) the minimum of the RMSE in the search space of HPs and 2) where (in HP space) to
acquire new observations so as to optimally improve the prediction for the minimum. Setting up a Bayesian
optimizer is not necessarily a straightforward task, since such a model comes with its own set of HPs.
In addition, the selection of HP combinations for acquiring new data is not done in an agnostic
way but influenced by where previous data have been acquired. Therefore, an efficient parallelization
strategy is not necessarily straightforward either. Using Bayesian optimization to improve the accuracy of ML
interatomic potentials is an active area of research, and we are currently undertaking efforts in this
direction. However, these sophisticated optimization strategies fall outside of the scope of the
present manuscript.

\begin{figure}[t]
\includegraphics[width=\columnwidth]{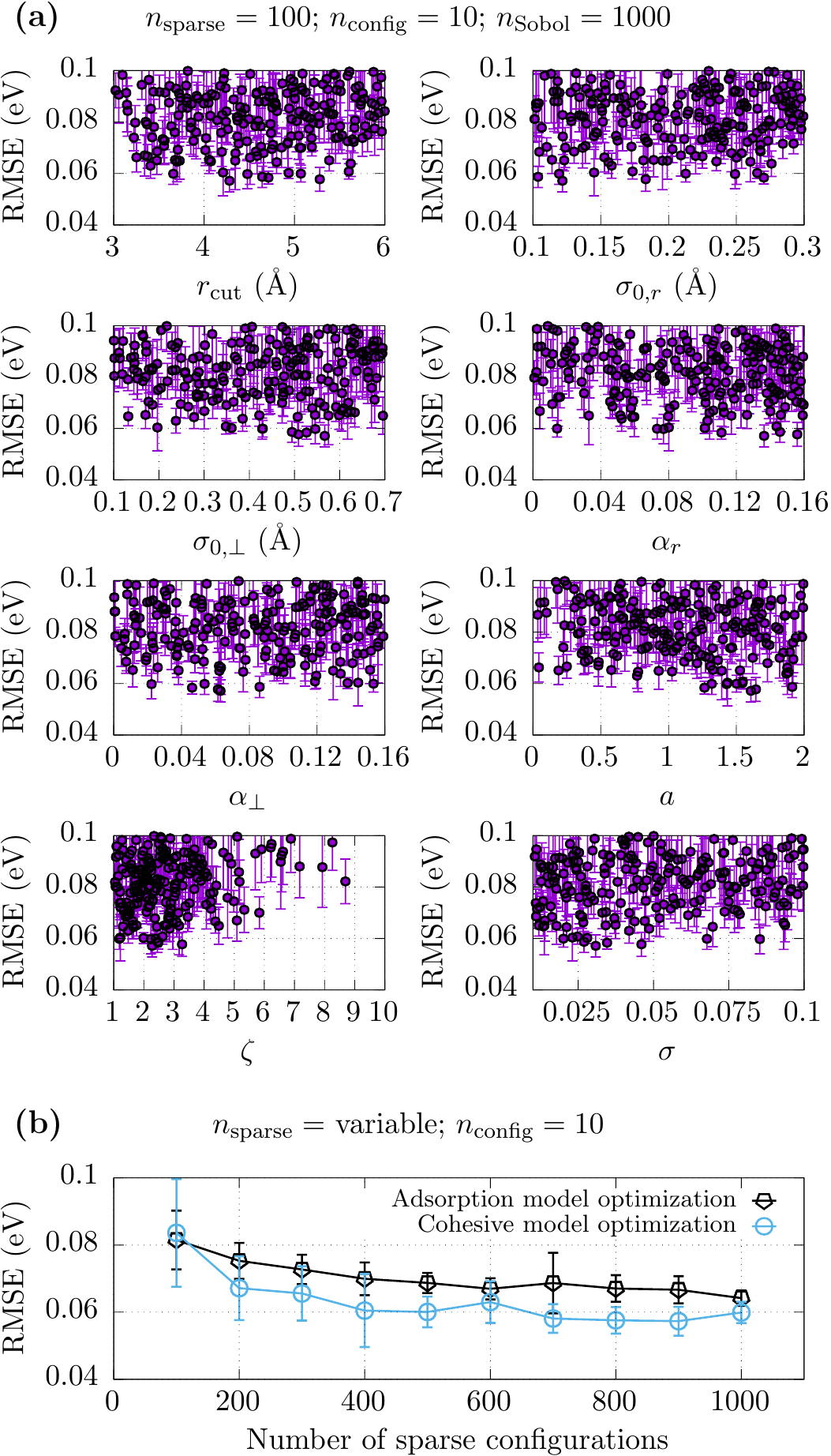}
\caption{(a) Average RMSE and standard deviation for cohesive energy GAPs
as a function of HPs. Each shown data point is the average among 10 models resulting from
10 randomly-chosen sparse sets. A total of 1000 Sobol vectors in 8D space are used, with each
dimension varying within the ranges shown in the figure. Therefore, a total of 10000 models
were trained. (b) With the optimal combination of HPs found in (a), we trained cohesive energy
GAPs with varying number of sparse-set configurations. Each data point shows the average RMSE
among 10 models (that is, the average for 10 different random selections of atomic structures in the
sparse set). The results are shown next to those from
\fig{13} for comparison.}
\label{17}
\end{figure}

Here, to optimize a GAP interatomic potential, we propose an intermediate solution between Bayesian optimization and the
random search carried out for our adsorption energy model in Sec.~\ref{16}: Sobol sequences~\cite{sobol_1976}. A Sobol
sequence is a series of points in an $N$-dimensional hypercube, where the points are chosen in sequence
so as to ``close holes'' in this space. These points fill space more homogeneously than randomly-chosen
or grid samples, but still without any knowledge of the objective function (therefore the Sobol
sequence is the same irrespective of which function is being sampled). A Sobol-sequence-based brute-force
search for optimal HPs is as trivial to parallelize as the random search from the previous section. In
\fig{17} we show the results of this analysis for a series of a-C GAP models trained from the same database
used in Sec.~\ref{18}. To have comparable results with respect to computational effort, we did not vary $n_\text{max}$
and $l_\text{max}$ in this search for optimal HPs, sticking to $n_\text{max} = l_\text{max} = 8$.

For the search in HP space, we used 1000 Sobol vectors~\cite{*[] [{. For the software implementation,
see github.com/naught101/sobol\_seq and people.sc.fsu.edu/{\textasciitilde}jburkardt/py\_src/sobol/sobol.html.}] sobol_1976}.
Each of the 1000 unique combinations of HPs was used to train 10 different GAP models. Each of these 10 models is built
with 100 sparse-set configurations, randomly chosen from the full training set. Figure~\ref{17}~(a) reports the
configurational average RMSE and error bars (standard deviation of the RMSE) as a function of HP. For these
cohesive energy GAPs, model performance as a function of HPs resembles our observations from \fig{14} for the
adsorption energy GAPs. In this case, optimal parameters are found around $\sigma_{0,r} = 0.12$~{\AA},
$\sigma_{0,\perp} = 0.55$~{\AA}, $\alpha_r = 0.14$, $\alpha_\perp = 0.06$, $a = 1.6$, $\zeta = 3$ and $r_\text{cut} = 4.3$~{\AA}.
In \fig{17}~(b) we use these HPs to repeat the same analysis we carried out in Sec.~\ref{18}. Our main observation is
that, while optimizing HPs directly on the same type of model (cohesive energy) for which these HPs are intended
improves model performance, it does so only marginally. The Sobol-based HP optimization based on reducing the RMSE for a cohesive energy
GAP produces optimal HPs which are similar to those obtained from an adsorption energy GAP,
and lead to final model improvement by $\sim 10\%$. Again, we highlight the idea that HPs are possibly transferable between
ML models that predict different properties, but that feed on the same type of structural data (namely, the atomic structure
as encoded via the SOAP descriptors). This raises the question of whether it is generally possible to use a computationally
cheap surrogate model (in our case the adsorption energy GAP) to optimize HPs for more computationally demanding
ML models (e.g., an interatomic potential or cohesive energy GAP). Based on the observations presented in this
paper, we speculate that this may indeed be the case, although further research is required in this direction.

\vfil

\section{Conclusions and outlook}

In this paper we have presented a modified form of the many-body atomic descriptor known
as SOAP~\cite{bartok_2013} and a series of mathematical and computational recipes for
its efficient evaluation. This type of descriptor is routinely used as essential input for
novel ML-based Gaussian approximation potentials~\cite{bartok_2010,bartok_2015} and other
ML models used to understand and predict the properties of solids and
molecules~\cite{caro_2018c,de_2016,jager_2018,himanen_2019}. While the primary objective of
this work was to improve the computational efficiency of SOAP calculations, the new
formulation also allows for a significant boost in accuracy. All in all, we expect that, at
fixed accuracy, the total speedup for practical applications will be between a factor of 10
and a factor of 20. This is a remarkable number because it means that ML-based
atomistic simulations can potentially be made one order of magnitude cheaper, bringing us one
(big) step closer to the realm of empirical interatomic potentials. ML-based simulations of
systems comprising up to one million atoms in the simulation box may now, or in the very
near future, become within reach.

\begin{acknowledgments}
The author is thankful to Volker Deringer and G\'abor Cs\'anyi, from the University
of Cambridge, Albert Bart\'ok, from the Science and Technology Facilities
Council, and Aki Morooka, formerly at Aalto University,
for very helpful discussions on SOAP descriptors and GAP models. We thank A. Bart\'ok again
for porting our implementation of the SOAP-like descriptor presented here into the QUIP code
(available as ``SOAP express''). The author also thanks
Dorothea Golze from Aalto University for her insights into basis set generation and code optimization. Funding
from the Academy of Finland under project 310574, computational resources from
CSC -- IT Center for Science and travel support from the HPC-Europa3 program are gratefully acknowledged.
\end{acknowledgments}

\end{document}